%% file: acl_latex.tex
\documentclass[11pt]{article}

\usepackage[]{acl}

\usepackage{times}
\usepackage{latexsym}

\usepackage[T1]{fontenc}
\usepackage[utf8]{inputenc}

\usepackage{microtype}

\usepackage{inconsolata}

\usepackage{graphicx}

\usepackage[table]{xcolor}
\usepackage{floatrow}
\usepackage{amssymb}
\usepackage[flushleft]{threeparttable}
\usepackage{booktabs}

\usepackage{lipsum}
\usepackage{pifont}% 
\usepackage{multirow} 
\usepackage{graphicx}
\usepackage{amsmath}
\usepackage[flushleft]{threeparttable}
\usepackage{subcaption}
\usepackage{colortbl}
\usepackage{mdframed}
\definecolor{lightorange}{HTML}{F59F05}
\definecolor{deeporange}{HTML}{EA5110}

\title{ProRank: Prompt Warmup via Reinforcement Learning for Small Language Models Reranking}

\author{%
  Xianming Li $^{2}$$^\dagger$\thanks{Work done during an internship at Mixedbread}, Aamir Shakir $^{1}$\thanks{Equal Contribution}, Rui Huang $^{1}$, \\
  \textbf{Tsz-fung Andrew Lee} $^{2}$, \textbf{Julius Lipp} $^{1}$, \textbf{Benjamin Clavi\'e} $^{1}$, \textbf{Jing Li} $^{2}$ \\
  $^1$ Mixedbread AI, Berlin, Germany\\
  $^2$ The Hong Kong Polytechnic University, Hong Kong SAR\\
  \texttt{\{sean,aamir,rui,julius,ben\}@mixedbread.ai} \\
  \texttt{xianming.li@connect.polyu.hk},\\
  \texttt{\{jing-amelia.li,tsz-fung-andrew.lee\}@polyu.edu.hk} \\
  \texttt{\includegraphics[width=0.33cm,height=0.32cm]{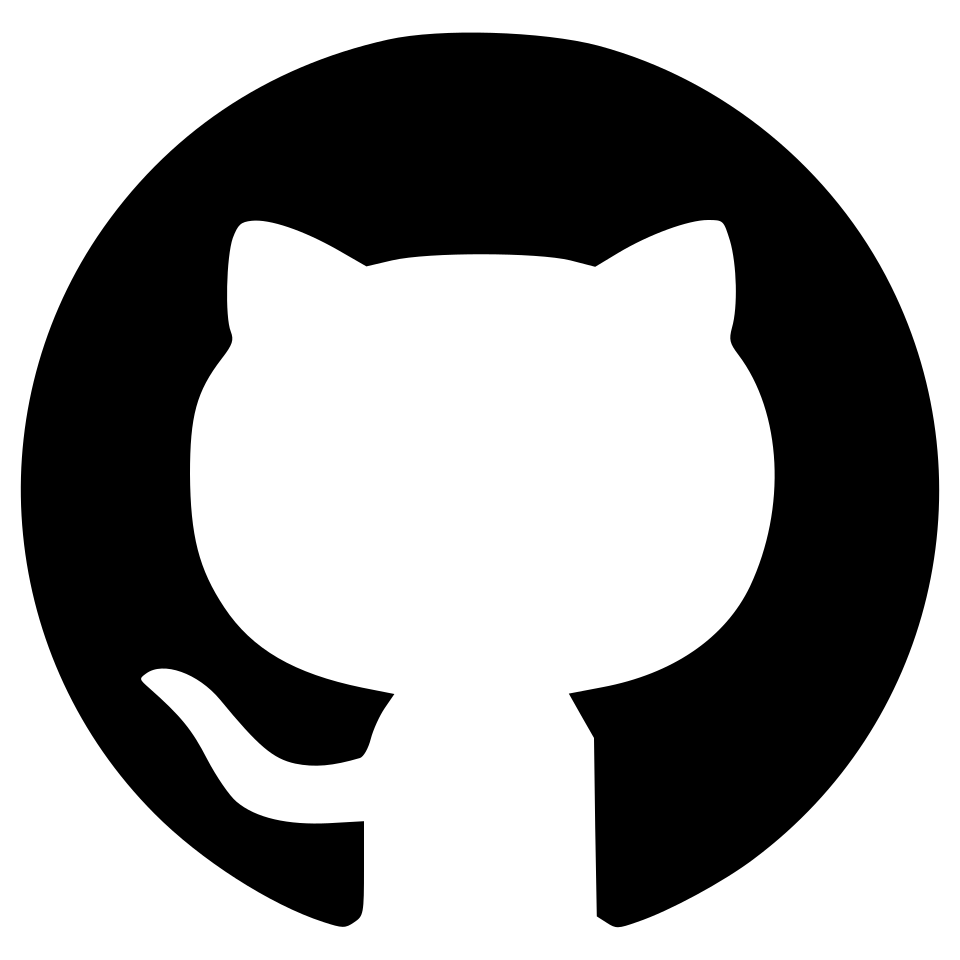} \href{https://github.com/mixedbread-ai/mxbai-rerank}{GitHub}} \thanks{ https://github.com/mixedbread-ai/mxbai-rerank} \ \ 
    \texttt{\includegraphics[width=0.33cm,height=0.32cm]{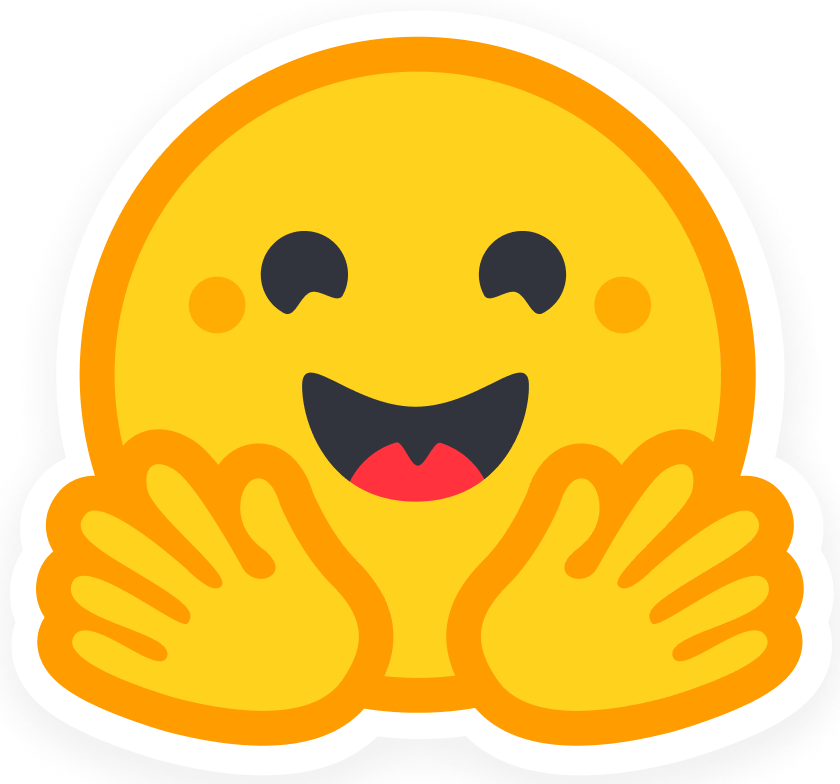} \href{https://hf.co/mixedbread-ai/mxbai-rerank-large-v2}{HuggingFace}} \thanks{ https://hf.co/mixedbread-ai/mxbai-rerank-large-v2} 
}

\begin{document}
\maketitle
\begin{abstract}
\input{abstract}
\end{abstract}

\section{Introduction}
\input{introduction}

\section{Preliminary Quantitative Analysis}
\label{sec:quantitative}
\input{quantitative}

\section{Methodology}
\label{sec:methodology}
\input{methodology}

\section{Experiment}
\label{sec:experiment}
\input{experiment}

\section{Related Work}
\input{related_work}

\section{Conclusion}
\input{conclusion}

\section*{Limitations}
A potential limitation observed in our experiments (Section \ref{sec:topk_retrieval}) is the challenge of handling noise in scenarios with very large top-$k$ (e.g., $k=5,000$) retrieval. 
To address this challenge, we recognize the need for enhanced semantic understanding capabilities that can better distinguish relevant results from noise.
For this limitation, our future work will pursue several promising directions: (1) developing more noise-robust reranking architectures, (2) investigating adaptive top-$k$ selection mechanisms that dynamically adjust based on query characteristics, and (3) extending ProRank's applicability to smaller, more efficient models and novel application domains.

\bibliography{custom}

\appendix

\input{appendix}

\end{document}

%% file: abstract.tex
Reranking is fundamental to information retrieval and retrieval-augmented generation, with recent Large Language Models (LLMs) significantly advancing reranking quality. 
Most current works rely on large-scale LLMs (>7B parameters), presenting high computational costs.
Small Language Models (SLMs) offer a promising alternative because of computational efficiency. 
However, our preliminary quantitative analysis reveals key limitations of SLMs: their representation space is narrow, leading to reduced expressiveness, and they struggle with understanding task prompts without fine-tuning. 
To address these issues, we introduce a novel two-stage training approach, \textbf{ProRank}, for SLM-based document reranking. 
We propose using reinforcement learning to improve the understanding of task prompts.
Additionally, we introduce fine-grained score learning to enhance representation expressiveness and further improve document reranking quality.
Extensive experiments suggest that ProRank consistently outperforms both the most advanced open-source and proprietary reranking models.
Notably, our $0.5$B ProRank even surpasses powerful LLM reranking models on BEIR benchmark, establishing that properly trained SLMs can achieve superior document reranking performance while maintaining computational efficiency.
% While recent advances with Large Language Models (LLMs) have significantly improved document reranking quality, 
% First, we propose a prompt warmup stage using reinforcement learning GRPO to steer SLMs to understand document ranking task prompts and improve their representation capabilities.
%% This stage enables the models to generate more accurate coarse-grained binary relevance scores for effective document reranking.
% Then, we continuously fine-tune the SLMs with a fine-grained score learning stage without introducing additional layers to further improve representation expresiveness and document reranking quality. 

%% file: introduction.tex
% TODO: a figure to draw the two limitations 
Document reranking is crucial in information retrieval and retrieval-augmented generation, which aims to reorder document lists initially retrieved by retrievers like BM25 based on query-document relevance \citep{zhu2023large}.
Very recently, Large Language Models (LLMs) have demonstrated remarkable performance in document reranking tasks \citep{ma2023zero, sun2023chatgpt, zhuang2023open, zhuang2024setwise, sun2024investigation, zhuang2025rank, weller2025rank1}, establishing a significant new direction in document reranking research.
Recent approaches have primarily employed LLMs through zero- or few-shot prompt engineering \citep{sun2023chatgpt, zhuang2023open, sun2024investigation, zhuang2025rank, weller2025rank1} to generate reordered document lists or coarse-grained binary relevance scores. 
However, they typically require larger LLMs (>7B parameters) to achieve promising results, posing challenges for real-world applications due to computational constraints. 
% as small language models (SLMs) struggle to understand task-specific prompts and generate high-quality ordered document lists in zero- or few-shot settings. 
% Large-scale LLMs also pose substantial challenges for real-world applications due to computational and resource constraints. 

To address this challenge, we explore using small language models (SLMs) for document reranking. 
Our investigation begins with a quantitative analysis. It suggests two key limitations of SLMs for document reranking: 1) SLMs are constrained by narrow representation spaces, impairing their capabilities for document reranking effectively; 
2) SLMs struggle to understand task prompts and generate proper coarse-grained binary relevance scores (0: irrelevant, 1: relevant) without proper fine-tuning. 
% These limitations hinder SLMs from being widely used for the document reranking task.

%% start here
To address these limitations of SLMs, we propose a novel two-stage approach, ProRank, for SLMs document reranking. 
% It unlocks the potential of SLMs for document reranking while producing interpretable fine-grained relevance scores. 
In the first stage, we employ reinforcement learning, specifically, GRPO (Group Relative Policy Optimization) \citep{shao2024deepseekmath}, to teach SLMs to understand the task prompt and produce properly formatted responses, i.e., coarse-grained binary relevance scores. 
The GRPO learning allows ProRank to incorporate various rewards into the output format and relevance accuracy, effectively teaching SLMs to generate binary relevance scores while maximizing relevance accuracy. 
However, these coarse-grained binary relevance scores are insufficient for high-quality document reranking, as they merely categorize documents as relevant (``1'') or irrelevant (``0'') without distinguishing relevance levels among documents in the same category.
%The coarse-grained binary relevance scores are too narrowly distributed to effectively differentiate relevance levels. 
%
Thus, we introduce the second stage -- fine-grained score learning. 
We employ an efficient method to generate fine-grained relevance scores by computing the relative scores between relevant (``1'') and irrelevant (``0'') token logit values from the model's last token logit outputs. 
It maintains computational efficiency by avoiding the introduction of additional layers while providing fine-grained scoring capabilities. 
In our further investigation, it also helps improve SLMs' representation expressiveness for the document reranking task.
We adopt the Cross-Encoder paradigm \citep{nogueira2019passage, shakir2024boost} to efficiently train ProRank, allowing for more effective and high-quality document reranking.

For a comprehensive evaluation, we extensively experiment on three benchmarks in various languages and applications: the widely used English BEIR benchmark \citep{thakur2021beir}, Chinese MTEB benchmark \citep{cmteb}, and Code Retrieval datasets \citep{li2024coir}.
Extensive results demonstrate that ProRank delivers high-quality document reranking performance. 
Notably, our $0.5$B ProRank outperforms even the $32$B fine-tuned LLM model on the English BEIR benchmark.
Through detailed empirical analysis, we show that ProRank effectively addresses the two key limitations of SLMs observed in the quantitative analysis.

In summary, our contributions are as follows:

~$\bullet$ Our quantitative analysis reveals two limitations of SLMs for the document reranking: 1) they exhibit a narrow representation space, limiting their expressive capabilities; 2) they struggle with understanding task prompts.

~$\bullet$ We propose a novel two-stage approach, ProRank, to effectively rerank documents with interpretable relevance scores, combining reinforcement learning for coarse-grained scoring with a fine-grained scoring.

~$\bullet$ Extensive evaluations demonstrate that ProRank achieves superior document reranking performance, with $0.5$B ProRank  outperforming strong baselines on various benchmarks. 
% the larger $32$B LLM reranking models.

%% file: quantitative.tex
Before introducing the proposed model -- ProRank, we conduct a preliminary quantitative analysis to evaluate how small language models (SLMs) perform on document reranking through zero-shot prompting.
We evaluate various popular SLMs using a consistent prompt (as follows) on the TRECCOVID test set ($66336$ samples) from the BEIR benchmark \cite{thakur2021beir}.
\input{prompt}
We visualize SLM's relative scores $\Delta = \text{TokenLogit}(1) - \text{TokenLogit}(0)$ in Figure \ref{fig:relative_score}, where $\text{TokenLogit}(1)$ and $\text{TokenLogit}(0)$ stand for the logit values corresponding to position of relevant (``1'') and irrelevant (``0'') tokens in the model's last token logits, respectively.

\input{figure_relative_score}

We observe two interesting phenomena: 1) \textbf{SLMs under 1B parameters exhibit a narrow representation space.} 
%While larger SLMs show improved representation capabilities, models like LLaMA and Gemma still demonstrate notably narrower representation spaces compared to Qwen. 
This narrow representation space could impair their effectiveness in document reranking tasks. 2) \textbf{SLMs demonstrate poor discriminative ability between ``relevant'' and ``irrelevant''}, evidenced by many relevant document points are below irrelevant points.
For phenomenon 1, it is related to the scale of models; larger models usually have more powerful representation capabilities.
For phenomenon 2, it is probably a common issue for SLMs to understand the task prompt without a prompt warmup stage. 
To verify this, we further investigate it. 
Specifically, we instruct these SLMs to generate binary relevance scores for given documents and queries. We measure both the format success rate (the ability to correctly generate  binary relevance scores ``0'' or ``1'') and accuracy (the correctness of relevance judgments compared to ground truth), as shown in Figure \ref{fig:format_success}.

\input{figure_format_success}

The results suggest that SLMs indeed struggle with understanding the ranking task prompt in zero-shot settings. 
Models like LLaMA completely fail at understanding the task, showing nearly $0\%$ in both format success rate and accuracy. 
Even more powerful models like Qwen show inconsistent performance, with format success rates varying unpredictably with model size.
Here, we also visualize the performance of the proposed model, ProRank, for a direct comparison. We will elaborate on it in the Section \ref{sec:task_prompt_understanding}.
% This extra evidence further verifies the observations made in phenomenon 2.

% Based on the quantitative analysis, the paper proposes ProRank to address two key issues identified from phenomena 1 and 2: (1) using reinforcement learning prompt warmup to better understand the document ranking task, and (2) employing fine-grained scoring learning to expand the representation space for the document ranking task.

%% file: prompt.tex
\begin{mdframed}[
    linewidth=2pt,
    linecolor=gray!50,
    backgroundcolor=gray!10,
    roundcorner=10pt,
    innertopmargin=10pt,
    innerbottommargin=10pt,
    innerrightmargin=15pt,
    innerleftmargin=15pt
]
    \label{frame:prompt}
    \textbf{Prompt}: 
    query: \textit{\{query\_placeholder\}}  \\
    document: \textit{\{document\_placeholder\}} \\
    You are a search relevance expert who evaluates how well documents match search queries. For each query-document pair, carefully analyze the semantic relationship between them, then provide your binary relevance judgment (0 for not relevant, 1 for relevant). Relevance:
\end{mdframed}

%% file: figure_relative_score.tex
\begin{figure*}[htbp]
    \centering
    
    % Top row: Llama and Gemma side by side
    \subfloat[LLaMA relative scores]{
        \includegraphics[width=0.48\linewidth]{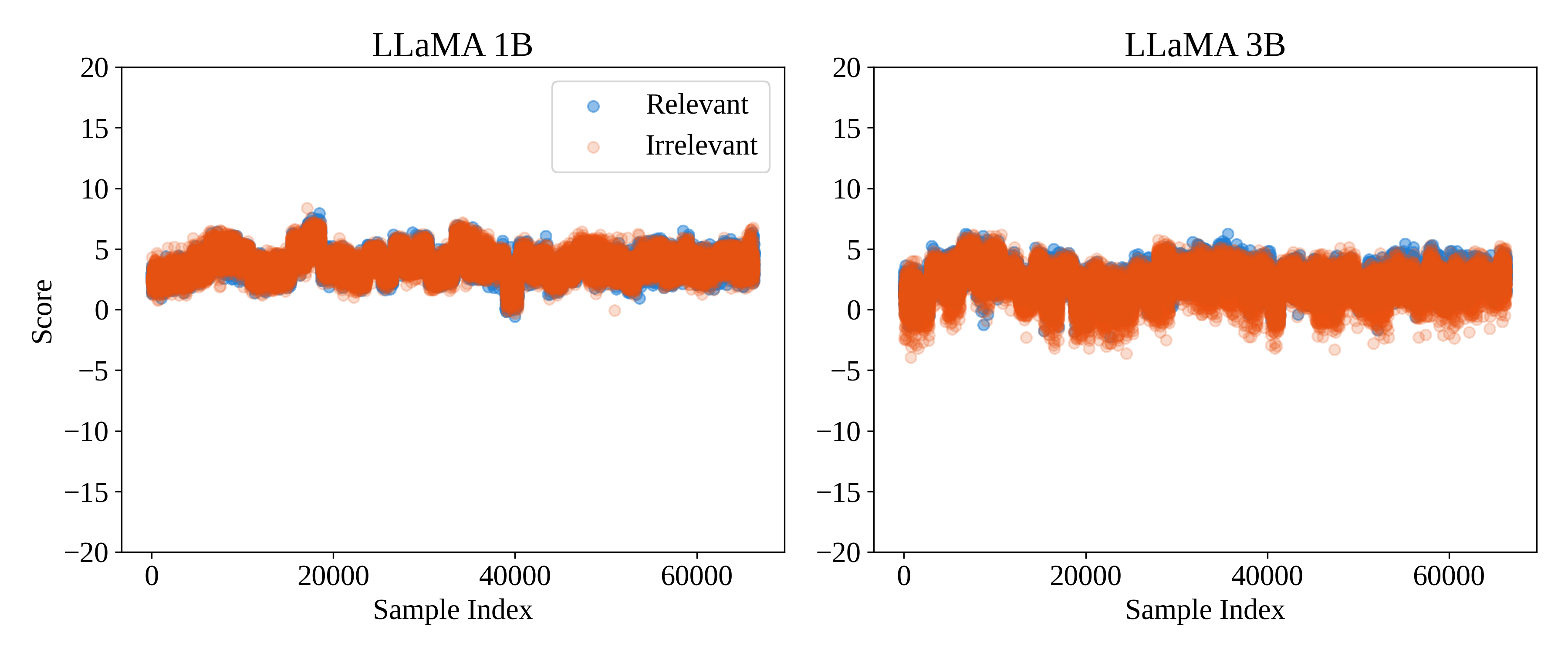}
        \label{fig:llama_logits}
    }
    \hfill
    \subfloat[Gemma relative scores]{
        \includegraphics[width=0.48\linewidth]{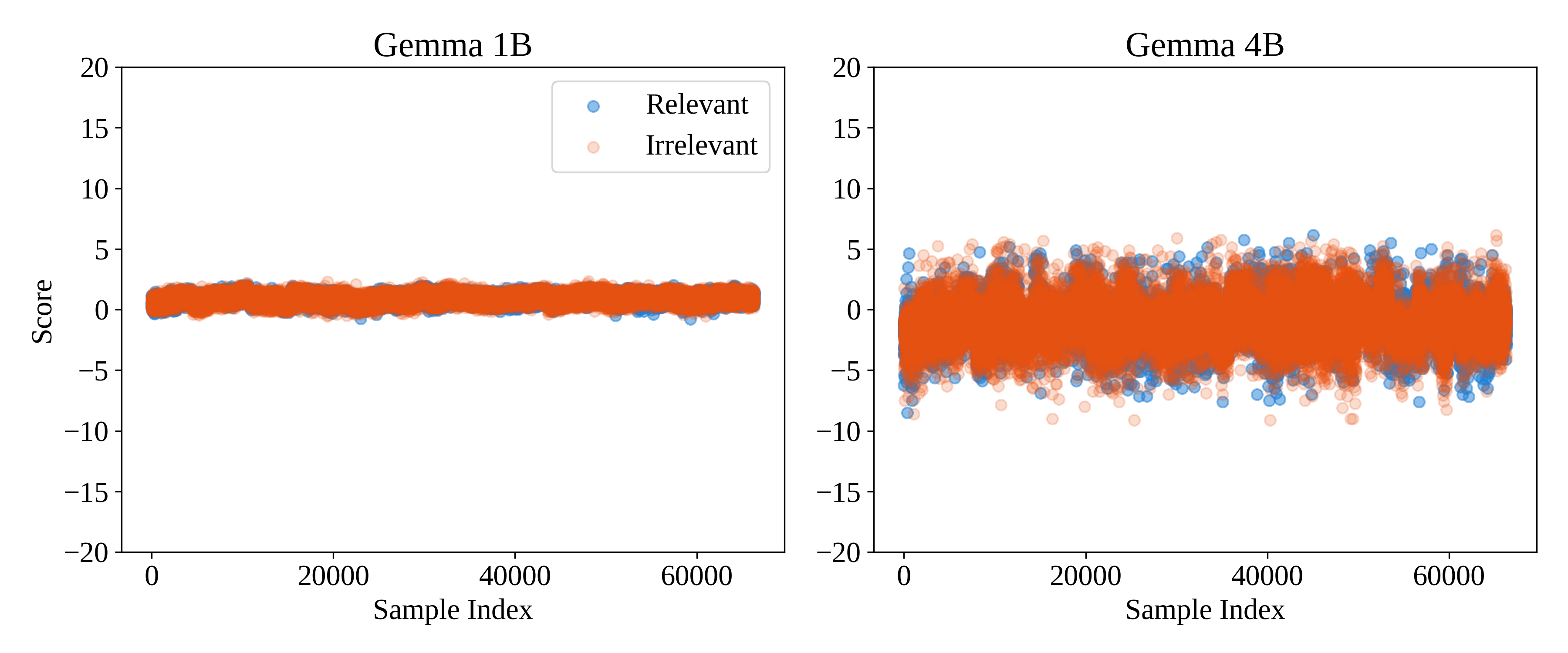}
        \label{fig:gemma_logits}
    }
    
    % \vspace{1em} % Add some vertical space
    
    % Bottom row: Qwen centered
    \subfloat[Qwen relative scores]{
        \includegraphics[width=\linewidth]{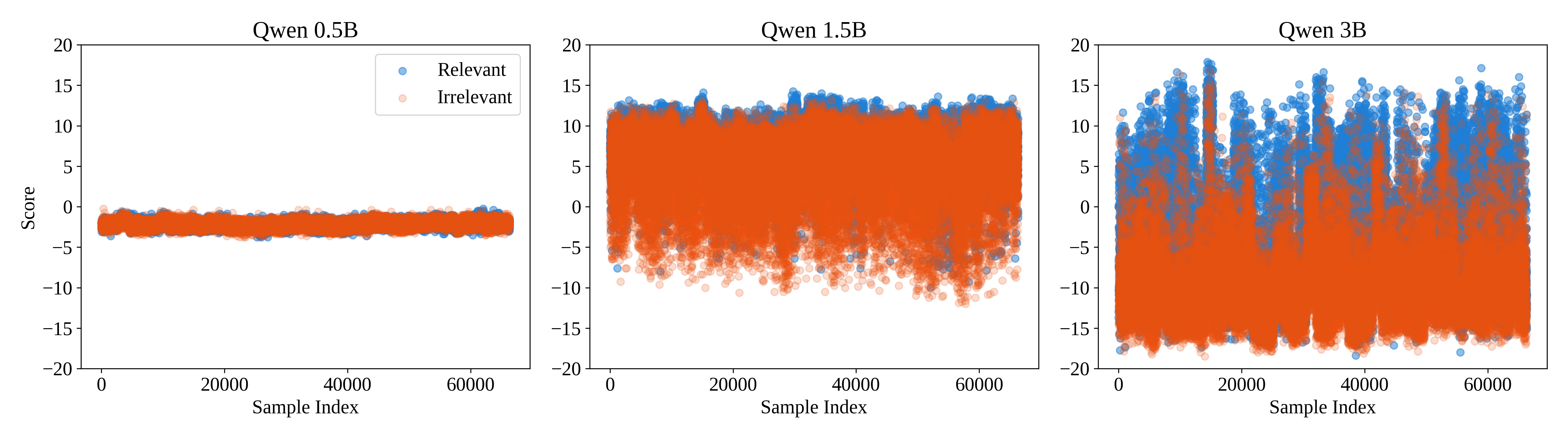}
        \label{fig:qwen_logits}
    }
    
    \caption{Visualization of relative scores between relevant (``1'') and irrelevant (``0'') tokens' logits on the TRECCOVID test set. For LLaMA, Gemma, and Qwen, we visualize the small language models (with size < 7B).}
    \label{fig:relative_score}
\end{figure*}

%% file: figure_format_success.tex
\begin{figure}[h!]
    \centering
    \includegraphics[width=\linewidth]{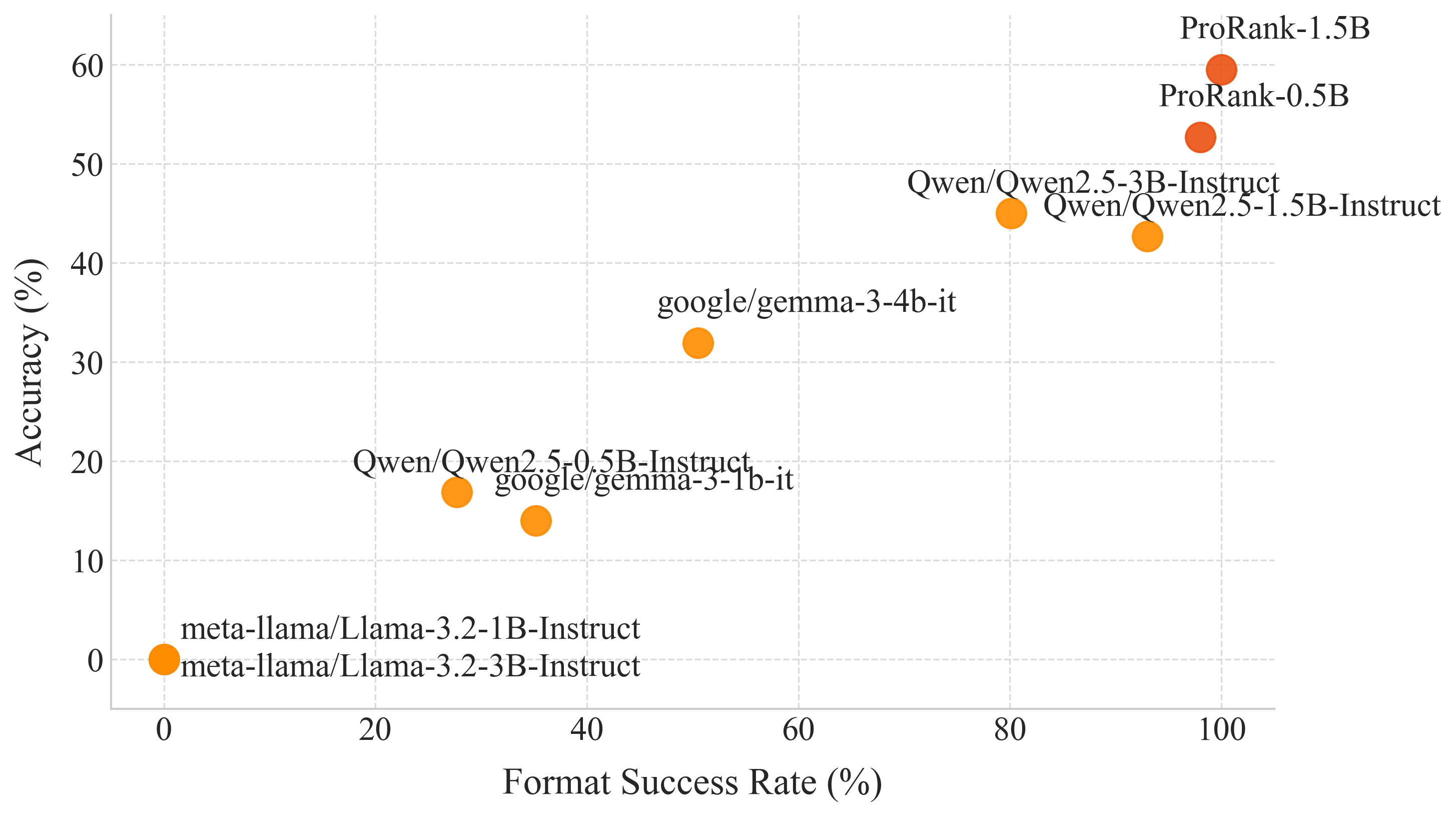}
    \caption{The accuracy ($y$-axis) and format success rate ($x$-axis) of SLMs in generating binary relevance scores on the TRECCOVID test set. The light orange colors indicate baselines' results. The dark orange colors denote the proposed ProRank's results.}
    \label{fig:format_success}
\end{figure}

%% file: methodology.tex
This section presents the proposed model, ProRank, with its generic framework illustrated in Figure \ref{fig:framework}.
We organize as follows: Section \ref{sec:prompt_warmup} introduces the first stage -- reinforcement learning prompt warmup, followed by Section \ref{sec:fine_grained_score_learning}, which details the second stage -- fine-grained score learning.

\input{figure_framework}

\subsection{Reinforcement Learning Prompt Warmup}
\label{sec:prompt_warmup}
The quantitative analysis reveals that SLMs face challenges in understanding task prompts and producing correctly binary relevance tokens without fine-tuning. 
Specifically, we employ GRPO \citep{shao2024deepseekmath} because it has been proven effective in optimizing multiple rewards, enabling SLMs to better learn task prompts and generate well-formatted output. The learning objective of GRPO is detailed in the appendix Section \ref{sec:appendix_grpo}.
We fine-tune ProRank using the prompt described in Section \ref{sec:quantitative}.
We optimize dual rewards to evaluate both format and relevance accuracy.
% to properly teach SLMs to generate well-formatted binary relevance tokens while ensuring relevance accuracy.
% To teach SLMs properly to generate responses with the specific format while improving relevance accuracy, the training incorporates a dual-aspect reward mechanism that evaluates both format and relevance accuracy. 
For the format reward, we define the reward function as follows:
\begin{equation}
    \label{eq:reward_1}
    r_1(o | prompt) = \begin{cases} 
        1 & \text{if}\ o \text{ is binary token} \\
        0 & \text{otherwise}
    \end{cases}
\end{equation}
We reward the model if it generates a response that follows the required binary format; otherwise, no reward will be given.
For relevance accuracy, we use the accuracy of the model's prediction as the reward function:
\begin{equation}
    \label{eq:reward_2}
    r_2(o | prompt) = \mathrm{accuracy}(o, y)
\end{equation}
where $y$ is the ground truth relevance label.
These dual rewards provide effective training signals throughout the reinforcement learning process, allowing SLMs to generate correctly formatted outputs while maximizing relevance accuracy for the document reranking.

\subsection{Fine-grained Score Learning}
\label{sec:fine_grained_score_learning}
While the reinforcement learning prompt warmup stage helps SLMs understand the task prompt well and produce correctly binary coarse-grained scores, i.e., ``1'' and ``0''. 
Its representation space is still narrow and lacks the necessary granularity for effective document reranking, making it difficult to rank documents with the same relevance score. 
% More fine-grained relevance scores are needed.

To address this limitation, we introduce a fine-grained score learning stage to enhance document reranking performance without introducing new layers. 
Specifically, ProRank computes a fine-grained relevance score by extracting and comparing logit values corresponding to binary tokens ``0'' and ``1'' from the last token's logits, as follows:
\begin{equation}
    \label{eq:fine_grained_score_learning}
    \begin{split}
    \Delta &= \text{TokenLogit}(1) - \text{TokenLogit}(0) \\
    & \text{TokenLogit}(t) = \mathbf{O}[{\text{token\_id}(t)}] \\
    & \mathbf{O} = \text{LLM}_{last\_pool}(prompt) \in \mathbb{R}^{V},
    \end{split}
\end{equation}
where $\text{token\_id}(t)$ maps an input token $t$ to its index in the model's vocabulary, $prompt$ represents the input prompt, $last\_pool$ denotes the last token pooling, $\mathbf{O}$ means the logit outputs at the last token position, and $V$ stands for the vocabulary size.
We extract logit outputs from the last token position since it encapsulates the input's complete semantics. This is because in the auto-regressive architecture of LLMs, the last token can attend to all previous tokens in the attention calculation. 
Additionally, the reinforcement learning prompt warmup stage ensures the model generates meaningful binary tokens with the last token logits.
By doing so, ProRank leverages the model's learned token logits to generate fine-grained relevance scores, without requiring additional parameters or architectural changes, making it both efficient and effective.

We train ProRank by minimizing the binary cross-entropy loss between the predicted fine-grained scores ($\hat{y}_i$) and the ground truth relevance labels ($y_i$), as follows:
\begin{equation}
    \label{eq:binary_cross_entropy_loss}
    \mathcal{L}_{\text{BCE}} = -\frac{1}{N}\sum_{i=1}^{N} \left[ y_i \log(\hat{y}_i) + (1 - y_i) \log(1 - \hat{y}_i) \right].
\end{equation}

%% file: figure_framework.tex
\begin{figure*}[h!]
    \centering
    \includegraphics[width=\linewidth]{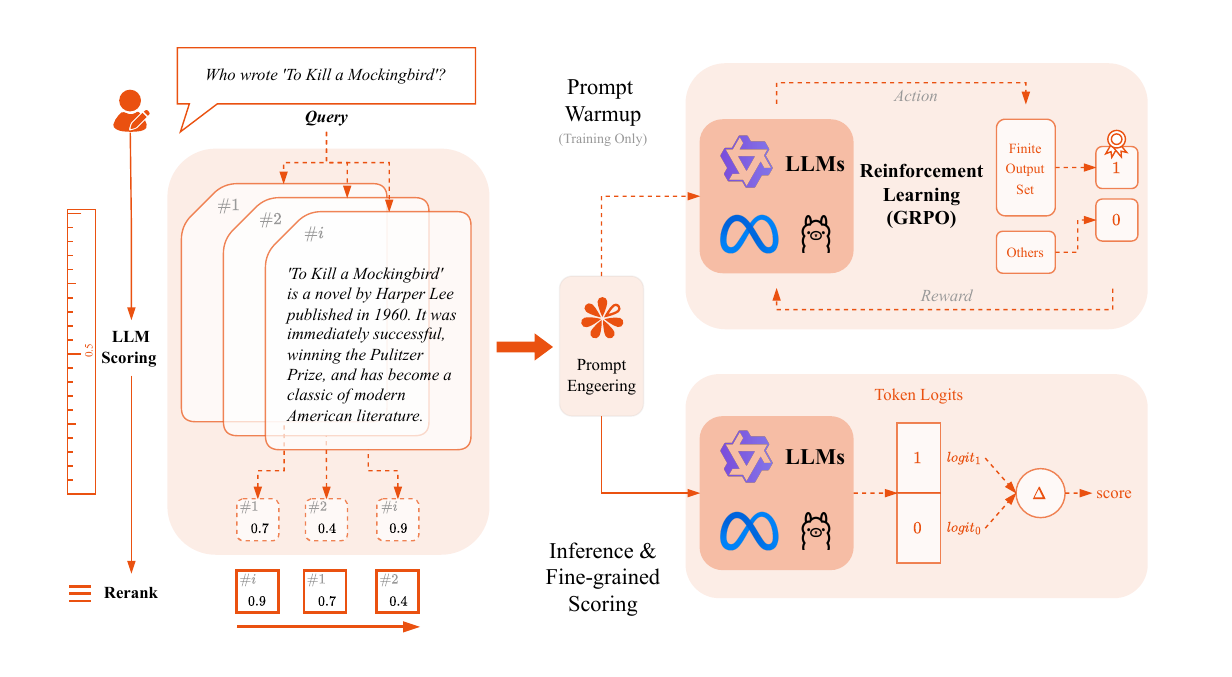}
    \vspace{-1.5em}
    \caption{The generic framework of the proposed model ProRank. There are two stages:  1) Prompt Warmup  (training only) steers LLM to produce coarse-grained binary relevance tokens (``1'' and ``0'') through reinforcement learning; 2) Fine-grained scoring stage computes relative logit scores to produce fine-grained scores for ranking.}
    \vspace{-1.5em}
    \label{fig:framework}
\end{figure*}

%% file: experiment.tex
\input{table_beir}

\subsection{Experiment Setup}
\label{sec:experiment_setup}
% Here, we elaborate on the experimental setup, including datasets, baselines, evaluation metrics, and implementation details.

\paragraph{Datasets}
We train ProRank using the bge-m3 dataset, following the popular reranking model \textit{BAAI/bge-reranker-v2-m3} \citep{chen2024bge}. 
This dataset is specifically curated for multilingual embedding and reranking tasks. 
For a comprehensive evaluation, we conduct experiments across multiple benchmarks spanning diverse languages and domains: (1) The BEIR benchmark \citep{thakur2021beir} for English reranking, where we utilize 14 datasets (details in Section \ref{sec:appendix_beir}). We further enhance our evaluation with additional English datasets, including news, robust04, and signal; (2) Chinese retrieval datasets from the C-MTEB benchmark \citep{cmteb} to assess Chinese-specific reranking capabilities, specifically Covid and DuReader; and (3) the COSQA dataset from the COIR benchmark \citep{li2024coir} to evaluate code reranking performance.
This diverse evaluation strategy enables us to thoroughly assess our model's generalizability across languages and domains, providing insights into its robustness and versatility in real-world applications.

\paragraph{Baselines}
To establish a comprehensive evaluation, we compare our model against popular state-of-the-art reranking models across different parameter scales. 
Our baselines include open-source models of varying sizes: 
\textit{mixedbread-ai/mxbai-rerank-large-v1} \citep{shakir2024boost} (mxbai), a BERT-scale model with 435M (0.4B) parameters known for its strong generalization capabilities;
\textit{BAAI/bge-reranker-v2-m3} \citep{chen2024bge} (bge-m3), a cutting-edge reranker with 568M (0.6B) parameters that represents the current state-of-the-art at the BERT scale;
and \textit{BAAI/bge-reranker-v2-gemma} \citep{chen2024bge} (bge-gemma), a larger 2.5B parameter SLM-based reranker that demonstrates the capabilities of models with significantly more parameters.
Additionally, we benchmark against two famous proprietary commercial reranking mdoels, including \textit{cohere-rerank-3.5} \citep{coherererank35} (cohere) and \textit{voyage-rerank-2} \citep{voyagererank2} (voyage), to position our approach within the broader landscape of available reranking solutions.

\paragraph{Evaluation Metric} We report Normalized Discounted Cumulative Gain at rank $10$ (NDCG@10), which measures the quality of ranking by considering both relevance and position, with higher scores indicating better ranking quality. 

\paragraph{Implementation Details}
We use the popular Qwen \citep{qwen2} as the backbone model and support two parameter scales: 0.5B and 1.5B. 
For efficient tuning, we employ LoRA \citep{hu2022lora} with hyperparameters lora\_r=$32$, lora\_alpha=$32$, and lora\_dropout=$0.1$ after a mild hyperparameter sweep.
The training process uses AdamW optimizer \citep{loshchilov2017decoupled} with an initial learning rate of $1e-4$ for both stages.
Following the common evaluation practice, we use BM25 as the first-stage retriever to retrieve initial document candidates. 
Following common practice, we retrieve the top $100$ documents for each query. 
All reranking models are then evaluated on the same retrieved documents for a fair comparison. 

\input{table_rest_main}

\subsection{Main Results}
\label{sec:main_results}

\paragraph{Results on BEIR Benchmark}
We first present our main results on the BEIR benchmark in Table \ref{table:beir}. 
% We observe that all rerankers improve the performance of the first-stage retriever, BM25, demonstrating the importance of reranking.
When comparing LLM-based models with BERT-variant models, we observe that LLM-based models generally achieve higher average scores across the benchmark. 
This performance advantage can be attributed to LLMs' larger parameter sizes and enhanced generalization capabilities, highlighting their effectiveness for document reranking.
Interestingly, we note that the BERT-variant model \textit{mixedbread-ai/mxbai-rerank-large-v1} still outperforms the state-of-the-art bge-gemma reranking model on five specific datasets. 
% We also observe that larger LLMs sometimes underperform smaller ones, for instance, Rank1-32B shows lower performance than MonoT5-3B on several metrics. 
This finding suggests considerable room for improvement in existing LLM-based reranking approaches.
% Another notable observation is that LLM-based reranking models perform comparably to commercial reranking solutions like \textit{cohere-rerank-3.5} and \textit{voyage-rerank-2}. 
% It further demonstrates LLMs' potential for reranking tasks, despite limited information about the proprietary commercial models.
% Overall, commercial reranking models generally outperform BERT-variant models on average scores across the benchmark.
Our 1.5B ProRank model with fine- and coarse-grained scoring demonstrates superior performance compared to baseline approaches, while even our smaller 0.5B variant remains competitive.
These results strongly validate the effectiveness of our proposed two-stage training methodology.
We also find that our ProRank with fine-grained score consistently outperforms the coarse-grained score, suggesting the importance of fine-grained scoring for effective document reranking.
% When compared against all baselines, our proposed fine-grained score model achieves superior performance across the BEIR benchmark, demonstrating the effectiveness of the proposed ProRank.
% Notably, our 0.5B and 1.5B ProRank outperform all baselines, including LLM-based reranking models.
% Our 1.5B ProRank further enhances performance, delivering significant improvements over both baselines and our 0.5B ProRank, with a substantial $1.92\%$ gain in average score.

\paragraph{Results on Additional Languages and Domains}
Table \ref{table:rest_main} presents the results on diverse datasets of English, Chinese, and code retrieval.
%, providing a comprehensive assessment of the cross-lingual and cross-domain capabilities of the proposed model. 
% The results suggest that all reranking models consistently outperform the BM25 first-stage retriever across all datasets, consistently proving the crucial role of reranking models.
Our proposed ProRank achieves superior performance compared to all baselines on nearly all datasets, with the only exception being the robust04 dataset, where \textit{mxbai} shows slightly better results. 
The performance improvements are particularly notable for our fine-grained score models, with the 0.5B and 1.5B ProRank delivering average gains of $1.93\%$ and $2.51\%$, respectively, over the powerful baseline--\textit{bge-gemma}.
% A key observation is that the fine-grained ProRank consistently outperforms its coarse-grained counterparts across all datasets, confirming the effectiveness of the proposed fine-grained scoring approach. 
% Interestingly, we observe divergent scaling behaviors between the two model types: coarse-grained models show performance degradation when scaling from 0.5B to 1.5B parameters, while fine-grained models demonstrate consistent improvement with increased scale. 
% This suggests that fine-grained scoring mechanisms are more robust to model scaling and can better leverage additional parameters. We will discuss more about the differences between the coarse-grained and fine-grained scores in the subsequent Section \ref{sec:granularity}. 
These findings further underscore the generalizability and effectiveness of ProRank in handling diverse languages and domains.

\subsection{Ablation Study}
\label{sec:ablation_study}
To better understand the contribution of each component in ProRank, we conduct a comprehensive ablation study.
First, we examine the impact of the fine-grained score learning stage. 
As shown in Table \ref{table:beir}, the fine-grained score learning significantly enhances reranking quality, confirming its critical role in our two-stage methodology.

To verify the importance of reinforcement learning prompt warmup, we train a 0.5B Qwen model without this initial stage. 
% Its result is only $35.06\%$ of NDCG@10 compared to $37.10\%$ NDCG@10 with reinforcement learning prompt warmup. 
It achieves a $2.04\%$ improvement when using reinforcement prompt warmup compared to without it, demonstrating that reinforcement learning prompt warmup is important for achieving better reranking performance. 

We also compare different fine-tuning strategies, specifically evaluating supervised fine-tuning (SFT) and the popular reinforcement learning algorithm--GRPO. 
The detailed comparison is in Appendix Section \ref{sec:appendix_sft_grpo}. The results indicate that using reinforcement learning for the first-stage prompt warmup achieves better performance than SFT, benefiting task prompt understanding.
% For this experiment, we also use 0.5B Qwen and maintain identical conditions across both methods.
%: the same training data (NLI subset from the bge-m3 dataset), hyperparameters, and random seed. 
% Our evaluation on the NFCorpus dataset reveals that while both methods achieve a perfect format success rate ($100\%$ success rate), GRPO significantly outperforms SFT with $62.27\%$ accuracy compared to SFT's $40.54\%$ accuracy. 
% This dramatic performance difference validates the superiority of GRPO for SLM reranking, as GRPO more effectively teaches SLMs to produce high-quality ranking scores.

\subsection{Discussion on Representation Capabilities}
\label{sec:representation_capability}
In the quantitative analysis, we observed that SLMs exhibit a narrow representation space without fine-tuning, resulting in limited representation expressiveness. Here, we visualize ProRank's relative scores in Figure \ref{fig:prorank_relative_score} to investigate this phenomenon.

For the 0.5B model, we observe a progressive widening of the representation space across training stages. It indicates that ProRank gradually learns to better distinguish between relevant and irrelevant documents, demonstrating improved representation expressiveness for the document reranking task.

For the 1.5B model, the pattern differs. The base Qwen 1.5B model already exhibits a relatively wide representation space, suggesting that the larger model possesses sufficient inherent expressiveness for the document reranking task. Consequently, the representation space does not undergo the same dramatic widening observed in the 0.5B ProRank across training stages. 
Nevertheless, the relative scores of relevant document points consistently shift upward and concentrate above irrelevant document points, suggesting the two-stage training process successfully teaches it to understand the task prompt, ultimately achieving high-quality document reranking through proper score calibration rather than representational expansion.

This evidence suggests that the proposed ProRank can effectively address the observed limitations in quantitative analysis.

\input{figure_prorank_relative_score}

\subsection{Discussion on Task Understanding}
\label{sec:task_prompt_understanding}
In the quantitative analysis, we found that SLMs struggle with understanding task prompts. 
To evaluate this, we visualize the format success rate of ProRank in Figure \ref{fig:format_success}. 
It shows that through the stage one prompt warmup learning, ProRank effectively learns to understand task prompts. 
Furthermore, as shown in Figure \ref{fig:prorank_relative_score}, both 0.5B and 1.5B ProRank demonstrate clear improvements in document ranking ability, with the relative scores of relevant document points increasing and clustering above those of irrelevant document points. This indicates that ProRank progressively develops a strong understanding of the document reranking task through the two-stage training process.

\input{figure_topk}

\subsection{Discussion on Top-$k$ Retrieval}
\label{sec:topk_retrieval}
Following \citet{jacob2024drowning}, we test different top-$k$ values (10, 100, 1,000, 5,000) on SciFact and FiQA datasets from BEIR to study how the number of retrieved documents affects reranking, as shown in Figure \ref{fig:topk}.
% We draw the following observations:
First, all reranking models demonstrate substantial performance gains when increasing the top-$k$ from $10$ to $100$ documents. 
It indicates that first-stage retrievers often miss highly relevant documents in the top $10$; therefore, having a larger candidate pool of up to $100$ is essential for effective reranking.
Second, our proposed ProRank consistently outperforms all baselines across all tested top-$k$ values on both datasets. 
% ProRank-1.5B, in particular, generally maintains the highest performance compared to baselines.
While the initial increase in top-$k$ is beneficial, we observe marginal effects beyond top-$k$=100 for both datasets.
% On the SciFact dataset, performance for most models tends to peak around top-$k$=1,000. Baselines and even the proposed ProRank shows a slight degradation when expanding to top-$k$=5,000.
% The FiQA dataset exhibits a similar trend; however, the benefits of increasing top-$k$ from 100 to 1000 appear less significant for some models, and the performance at top-$k$=5,000 does not witness a performance drop compared to SciFact. 
This evidence suggests that while increasing the number of candidates by expanding the top-$k$ can improve ranking quality, it also introduces a higher volume of non-relevant documents (noise). 
This noise can challenge the reranking models, potentially diluting the impact of truly relevant documents, especially at very large top-$k$ values like 5,000. 
Our ProRank, while generally robust, also shows sensitivity to noise at higher top-$k$ values, highlighting an ongoing challenge in document reranking.

% These findings emphasize the importance of carefully selecting the top-$k$ value for different datasets or domains. 
% While a larger top-$k$ can improve potential recall, it comes with increased computational costs for the reranker and the risk of performance degradation due to noise. 

\subsection{More Discussions}
To gain more insights, we present a detailed case study in Appendix Section \ref{sec:appendix_case_study}, analyze the performance differences between coarse- and fine-grained scoring in Appendix Section \ref{sec:appendix_granularity}, compare ProRank with cutting-edge LLMs for document reranking task in Appendix Section \ref{sec:appendix_cutting_edge}, and discuss the efficiency of ProRank in Appendix Section \ref{sec:appendix_efficiency}.

%% file: table_beir.tex
\begin{table*}[htbp]
\setlength\tabcolsep{2pt}
\small
\centering
\begin{threeparttable}
\caption{Results on BEIR benchmark, with NDCG@10 as the main metric. The orange color indicates the best values. $\dagger$ indicates results are retrieved from \citep{zhuang2025rank}. 
% $\ddagger$ means results are from \citep{weller2025rank1}. 
Our results are the average of five runs. Our 1.5B ProRank achieves the best result with significant performance gains compared to baselines on average ($p < 1\%$).
}
\label{table:beir}
\begin{tabular}{lcccccccccccccccccc}
\toprule
Model & Params & SF & NFC  & TC & AA & FQA & QRA & DBP & TCH & SD & FVR & CFV & HQA & MM & NQ & Avg. \\
\midrule
\midrule
\multicolumn{17}{c}{First-stage Retriever} \\
BM25 & $-$ & $67.89$ & $33.75$ & $59.47$ & $29.99$ & $23.61$ & $78.86$ & $31.80$ & $44.22$ & $14.91$ & $65.13$ & $16.51$ & $63.30$ & $22.84$ & $30.55$ & $41.63$ \\
\midrule
\midrule
\multicolumn{17}{c}{BERT-variant Reranker} \\
mxbai & $0.4$B & $74.83$ & \cellcolor{lightorange}$39.03$ & \cellcolor{lightorange}$85.33$ & $12.06$ & $40.46$ & $73.90$ & $44.66$ & $35.56$ & $18.89$ & $78.92$ & $23.92$ & $70.85$ & $36.31$ & $55.69$ & $49.32$ \\
bge-m3 & $0.6$B & $74.5$ & $35.89$ & $77.74$ & $51.98$ & $39.50$ & $89.02$ & $44.05$ & $35.14$ & $14.91$ & $83.45$ & $29.19$ & $79.5$ & $41.73$ & $58.56$ & $53.94$ \\
\midrule
\midrule
\multicolumn{17}{c}{LLM-based Reranker} \\
MonoT5$^\dagger$ & $3$B & $76.10$ & $37.80$ & $79.60$ & $42.50$ & $46.50$ & $-$ & $44.50$ & $30.70$ & $19.30$ & $-$ & $25.40$ & $-$ & $-$ & $-$ & $-$ \\
RankLlama$^\dagger$ & $7$B & $71.10$ & $27.00$ & $80.20$ & $54.40$ & $42.10$ & $-$ & $43.70$ & $41.40$ & $16.60$ & $-$ & $23.20$ & $-$ & $-$ & $-$ & $-$ \\
RankLlama$^\dagger$ & $13$B & $72.70$ & $28.10$ & $80.80$ & $49.30$ & $44.10$ & $-$ & $44.90$ & \cellcolor{lightorange}$39.20$ & $18.10$ & $-$ & $24.50$ & $-$ & $-$ & $-$ & $-$ \\
% Rank1$^\ddagger$ & $7$B & $77.20$ & $36.20$ & $81.90$ & $42.80$ & $39.50$ & $-$ & $38.90$ & $22.80$ & $17.20$ & $-$ & $15.00$ & $-$ & $-$ & $-$ & $-$ \\
% Rank1$^\ddagger$ & $14$B & $77.00$ & $35.80$ & $78.20$ & $45.30$ & $37.90$ & $-$ & $37.40$ & $27.10$ & $17.90$ & $-$ & $16.20$ & $-$ & $-$ & $-$ & $-$ \\
% Rank1$^\ddagger$ & $32$B & $76.80$ & $36.90$ & $81.90$ & $57.60$ & $41.80$ & $-$ & $40.70$ & $19.90$ & $19.60$ & $-$ & $15.80$ & $-$ & $-$ & $-$ & $-$ \\
bge-gemma & $2.5$B & $78.11$ & $38.35$ & $78.13$ & $54.6$ & $41.03$ & $89.23$ & $44.27$ & $32.93$ & \cellcolor{lightorange}$19.66$ & $84.39$ & $31.55$ & \cellcolor{lightorange}$80.21$ & \cellcolor{lightorange}$41.83$ & $60.95$ & $55.37$ \\
\midrule
\midrule
\multicolumn{17}{c}{Proprietary Reranker} \\
cohere & $-$ & $76.58$ & $34.65$ & $79.5$ & $59.90$ & $43.40$ & $87.12$ & $45.12$ & $37.20$ & $18.70$ & $86.20$ & $30.10$ & $77.30$ & $40.60$ & $59.09$ & $55.39$ \\
voyage & $-$ &  $76.25$ & $35.50$ & $79.9$ & $61.5$ & \cellcolor{lightorange}$46.7$ & $87.6$ & $43.5$ & $28.44$ & $18.2$ & $85.8$ & $21.4$ & $78.02$ & $41.57$ & $59.12$ & $54.54$ \\
\midrule
\midrule
\multicolumn{17}{c}{ \textit{Coarse-grained Reranker}} \\ 
ProRank & 0.5B & $76.16$ & $37.78$ & $77.21$ & $59.21$ & $40.51$ & $87.51$ & $42.93$ &  $31.35$ & $17.60$ & $88.77$ & $31.98$ & $78.51$ & $37.26$ & $56.89$ & $54.55$ \\
ProRank & 1.5B & $79.76$ & $35.90$ &  $77.70$ & $67.89$ &  $42.25$ &  $88.95$ & $40.97$ &  $28.00$ &  $17.66$ & $89.87$ & $36.40$ & $79.23$ & $35.62$ & $59.95$ & $55.73$ \\

\midrule

\multicolumn{17}{c}{ \textit{Fine-grained Reranker}} \\
ProRank & 0.5B & $77.15$ & $37.10$ & $79.06$ & $57.82$ & $41.09$ & $88.61$ & $44.66$ & $34.58$ & $17.02$ & $88.55$ & $33.02$ & $79.14$ & $41.25$ & $58.89$ & $55.57$ \\
ProRank & 1.5B & \cellcolor{lightorange}$80.87$ & $37.06$ & $80.01$ & \cellcolor{lightorange}$67.97$ & $44.41$ & \cellcolor{lightorange}$89.26$ & \cellcolor{lightorange}$45.63$ & $32.06$ & $17.83$ & \cellcolor{lightorange}$89.86$ & \cellcolor{lightorange}$37.12$ & $79.66$ & $41.75$ & \cellcolor{lightorange}$61.43$ & \cellcolor{lightorange}$57.49$ \\
\bottomrule
\end{tabular}
\end{threeparttable}
\end{table*}

%% file: table_rest_main.tex
\begin{table*}[htbp]
    \small
    \centering
    \begin{threeparttable}
    \caption{Results on other English, Chinese, and code retrieval datasets, with NDCG@10 as the primary metric. The orange color indicates the best values across all models.}
    \label{table:rest_main}
    \begin{tabular}{lcccccccccc}
    \toprule
    \multirow{2}{*}{Model} & \multirow{2}{*}{Params} & news  &	robust04  & signal & Covid  & DuReader & COSQA & \multirow{2}{*}{Avg.} \\
     & & EN &	EN & EN & CN & CN & Code &  \\
    \midrule
    \midrule
    \multicolumn{9}{c}{ \textit{First-stage Retriever}} \\
    BM25 & $-$ & $39.52$ & $40.70$ & $33.05$ & $76.10$ & $53.39$ & $21.79$ & $44.09$  \\
    \midrule
    \midrule
    mxbai & $0.4$B & $39.52$ & \cellcolor{lightorange}$56.38$ & $31.98$ & $78.60$ & $66.46$ & $30.72$ & $50.61$ \\
    bge-m3 & $0.6$B & $43.21$ & $50.24$ & $30.66$ & $85.94$ & $77.72$ & $24.86$ & $52.11$ \\
    bge-gemma & $2.5$B & $47.22$ & $51.94$ & $32.84$ & $78.55$ & $78.44$ & $31.51$ & $53.42$ \\
    \midrule
    \midrule
    \multicolumn{9}{c}{ \textit{Coarse-grained Reranker}} \\ 
    ProRank & $0.5$B & $47.54$ & $51.45$ & $33.85$ & $85.44$ & $76.52$ & $30.31$ & $54.19$ \\
    ProRank & $1.5$B & $47.28$ & $53.36$ & $30.96$ & $85.55$ & $75.46$ & $29.35$ & $53.66$ \\
    \midrule
    \multicolumn{9}{c}{ \textit{Fine-grained Reranker}} \\ 
    ProRank & $0.5$B & $46.57$ & $52.25$ & \cellcolor{lightorange}$34.12$ & $89.37$ & $78.03$ & $31.73$ & $55.35$ \\
    ProRank & $1.5$B & \cellcolor{lightorange}$49.06$ & $54.32$ & $31.85$ & \cellcolor{lightorange}$89.78$ & \cellcolor{lightorange}$78.54$ & \cellcolor{lightorange}$32.05$ & \cellcolor{lightorange}$55.93$ \\
    \bottomrule
\end{tabular}
\end{threeparttable}
\end{table*}

%% file: figure_prorank_relative_score.tex
\begin{figure*}[h!]
    \centering
    \includegraphics[width=\linewidth]{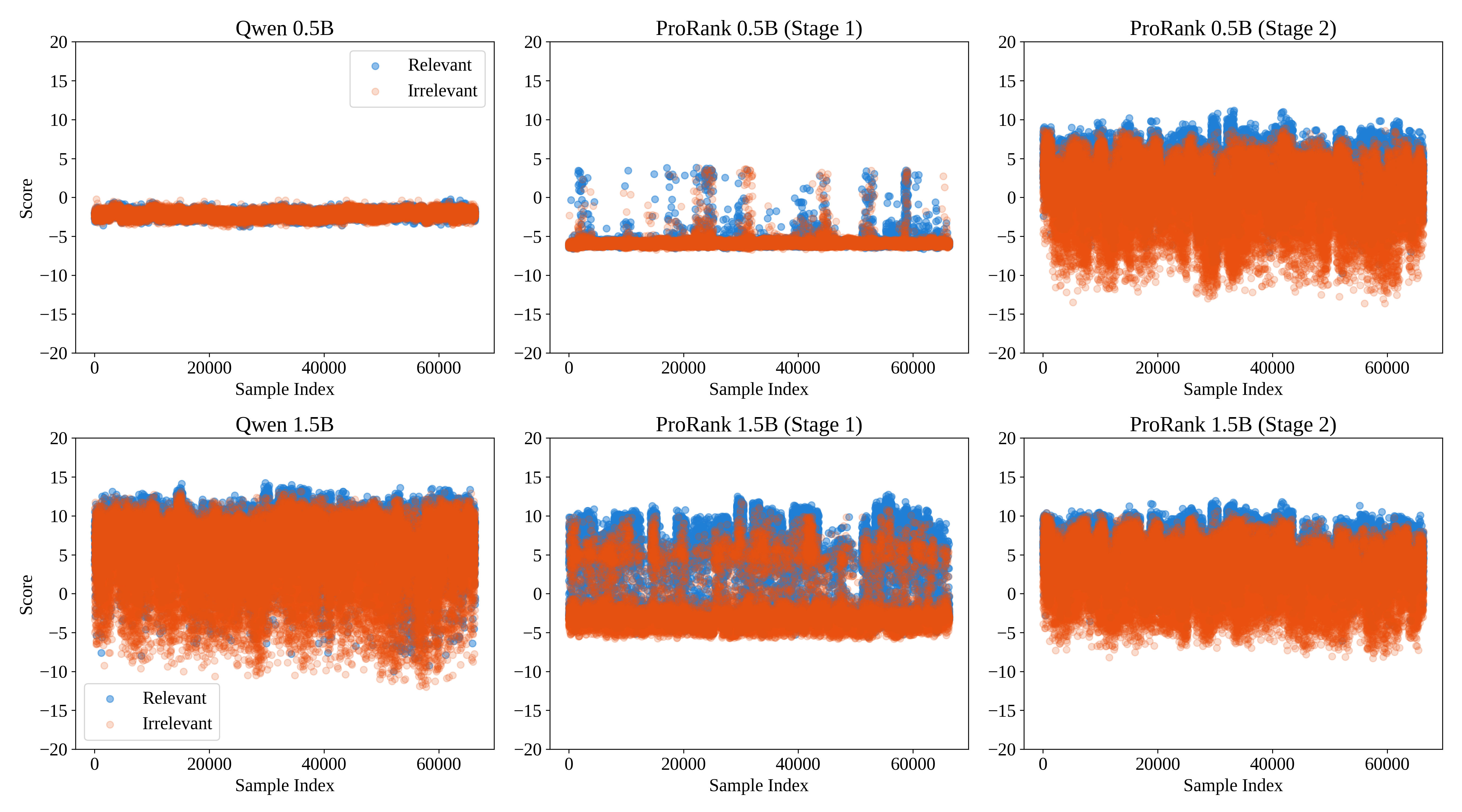}
    \vspace{-1.5em}
    \caption{Visualization of relative scores between relevant (``1'') and irrelevant (``0'') tokens' logits on the TRECCOVID test set for Qwen and ProRank.}
    \label{fig:prorank_relative_score}
\end{figure*}

%% file: figure_topk.tex
\begin{figure*}
     \centering
     \begin{subfigure}[b]{0.49\textwidth}
         \centering
         \includegraphics[width=\textwidth]{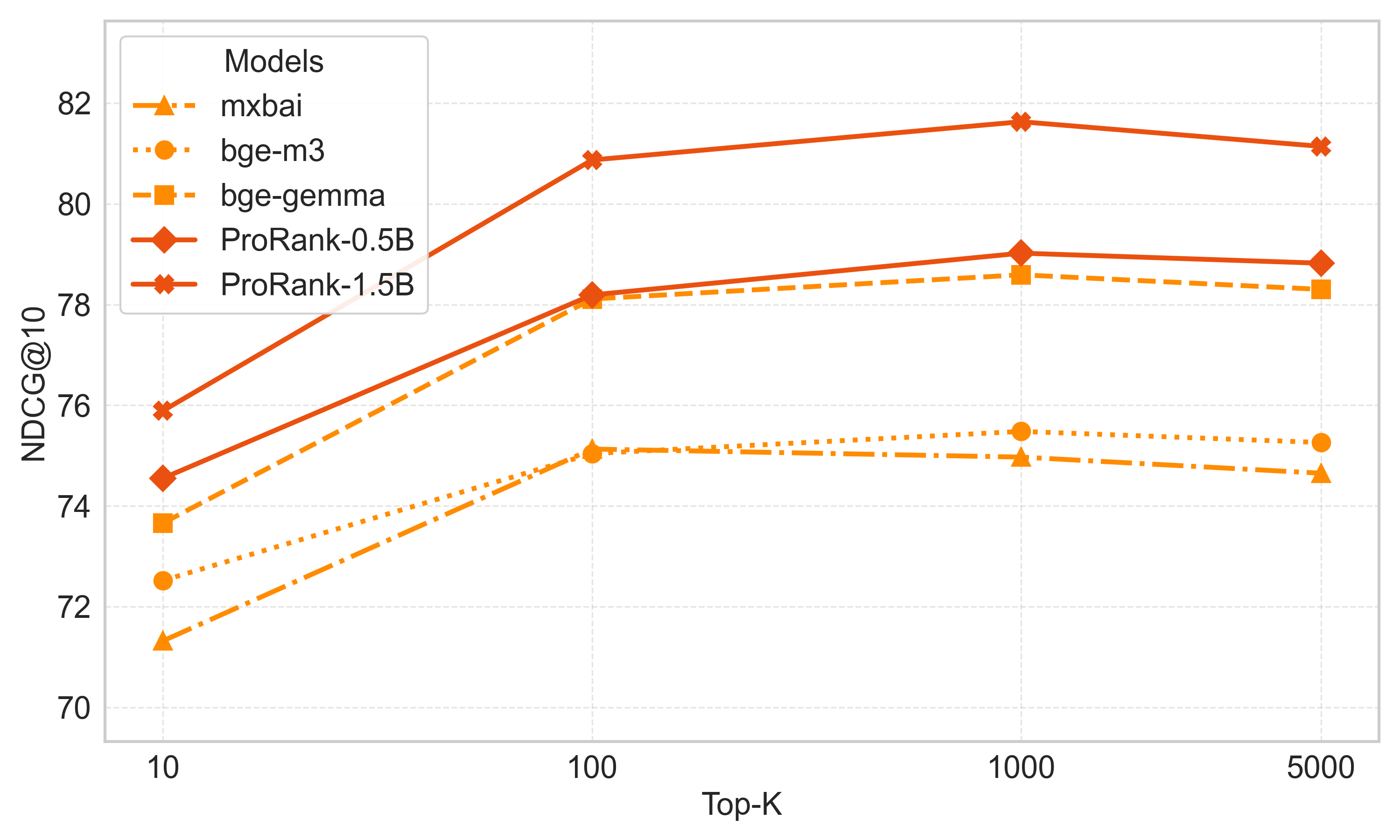}
         \caption{Top-$k$ Retrieval  on SciFact}
         \label{fig:topk_scifact}
     \end{subfigure}
     \hfill
     \begin{subfigure}[b]{0.49\textwidth}
         \centering
         \includegraphics[width=\textwidth]{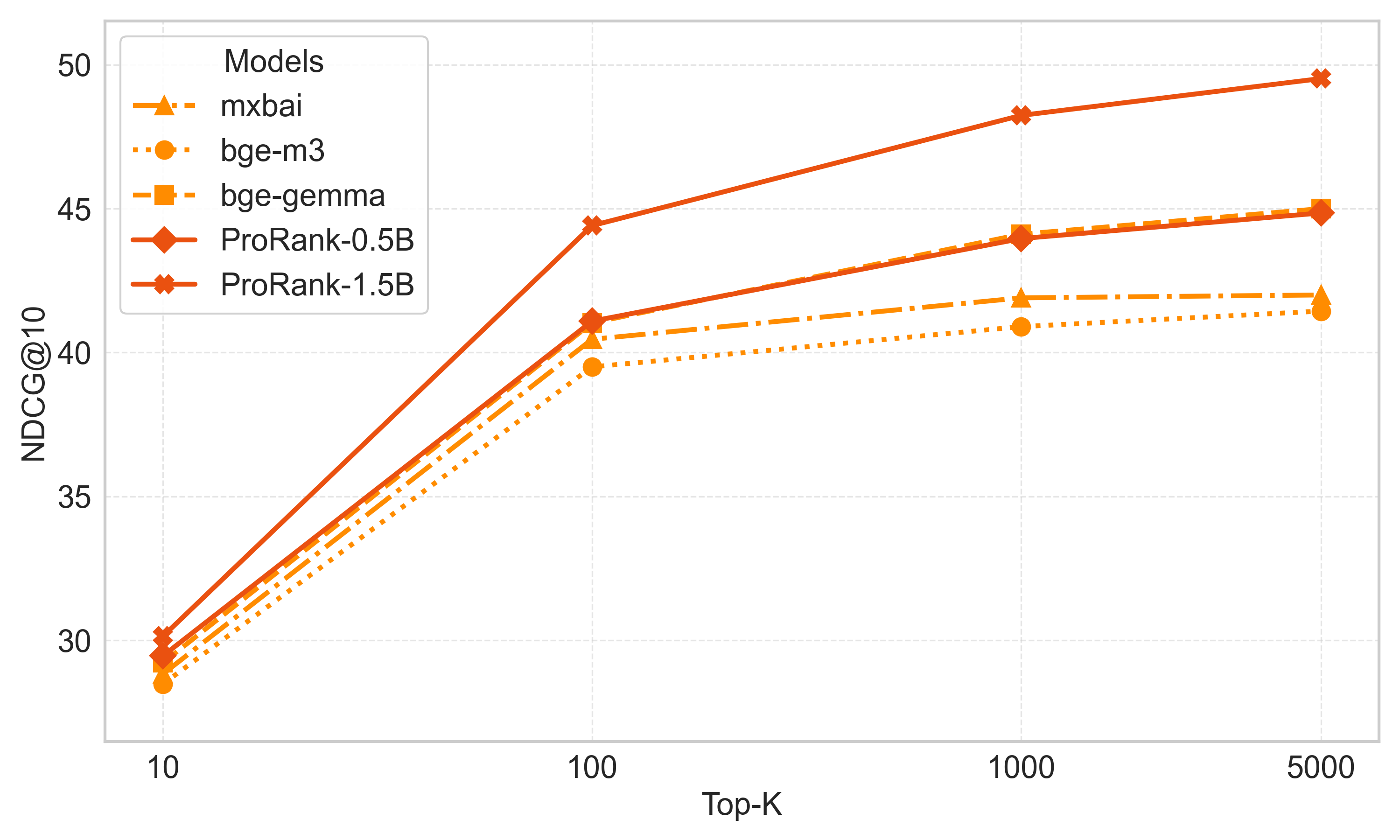}
         \caption{Top-$k$ Retrieval  on FiQA}
         \label{fig:topk_fiqa}
     \end{subfigure}
     \caption{Performance comparison of our fine-grained reranker across different numbers of retrieved documents (top-$k$) on the SciFact and FiQA dataset.}
    \label{fig:topk}
\end{figure*}

%% file: related_work.tex
This work is in line with document reranking. 
Reranking has traditionally been implemented using BERT-based models. 
\citet{nogueira2019passage} pioneered this approach with BERT for passage reranking, establishing a framework where query-document pairs are encoded together to generate relevance scores. 
These models, often referred to as Cross-Encoders, have become the standard for traditional reranking systems. 
Recent advancements, such as \textit{mixedbread-ai/mxbai-rerank} \citep{shakir2024boost} and \textit{bge-reranker} \citep{chen2024bge}, optimize the Cross-Encoder architecture for improved generalization across diverse retrieval tasks. 

The emergence of large language models (LLMs) has created new paradigms for reranking. 
Zero-shot approaches leverage LLMs' inherent capabilities without task-specific training \citep{zhuang2023open, ma2023zero}. 
\citet{sun2023chatgpt} demonstrated that LLMs like ChatGPT can effectively rerank documents through carefully designed prompting engineering. 
\citet{zhuang2024setwise} further advanced this direction with setwise ranking approaches that consider multiple documents simultaneously. 
To further improve the performance of LLMs for reranking, recent work has focused on fine-tuning LLMs for reranking. \textit{Rank1} \citep{weller2025rank1} introduced techniques to optimize compute efficiency during inference time. 
\citet{zhuang2025rank} proposed \textit{Rank-R1}, which enhances LLM reasoning capabilities for document reranking through reinforcement learning. 
Commercial models like \textit{cohere-rerank-3.5} \citep{coherererank35} and \textit{voyage-rerank-2} \citep{voyagererank2} offer powerful document reranking capabilities and strong generalization ability.

Despite these advancements, current reranking approaches still have limitations.
Zero-shot methods require larger LLMs to function effectively, but these models are challenging to interpret and computationally expensive to utilize in practice \citep{zhu2023large}.
Our proposed ProRank differs from existing work by enabling small language models (SLMs) to perform high-quality reranking while producing interpretable, fine-grained relevance scores, addressing significant limitations in current LLM-based reranking systems.

%% file: conclusion.tex
In this paper, we present ProRank, a novel two-stage approach for document reranking using Small Language Models (SLMs). 
Through a preliminary quantitative analysis, we identify two key limitations of SLMs in document reranking: inadequate task prompt understanding and constrained representation space.
To address the first limitation, ProRank uses reinforcement learning for prompt warmup learning.
Additionally, ProRank employs a fine-grained scoring mechanism to expand the representation space, enhancing representation expressiveness and delivering superior document reranking performance.
% The core innovation of ProRank is the combination of GRPO for coarse-grained score generation, followed by a fine-grained scoring learning. 
Comprehensive experiments on various domains and languages have verified the effectiveness and generalizability of ProRank. 
% Notably, our ProRank 0.5B model outperforms state-of-the-art baselines, including larger LLM-based and proprietary rerankers, while our ProRank 1.5B model further enhances performance with a substantial gain.
By enabling SLMs to achieve competitive performance, ProRank makes high-quality reranking more accessible for resource-constrained environments.

%% file: appendix.tex
\section{GRPO Learning}
\label{sec:appendix_grpo}
We use reinforcement learning in the first stage of ProRank to teach small language models (SLMs) to effectively understand document ranking task prompts and produce proper signal tokens for document reranking. 
Specifically, we adopt GRPO (Group Relative Policy Optimization) \citep{shao2024deepseekmath}, which is a popular reinforcement learning algorithm that has been proven effective for optimizing multiple rewards simultaneously \citep{shao2024deepseekmath}.
The GRPO learning objective is as follows:
\begin{equation}
    \begin{split}
        J_{\mathrm{GRPO}}(\theta)
    &= \mathbb{E}\left [q \sim P(Q), \{o_i\}_{i=1}^G \sim \pi_{\theta_{\mathrm{old}}}(O |q)\right] \\
    &
    \frac{1}{G} \sum_{i=1}^G
    \Bigl(
    \min\!\bigl(
    \frac{\pi_\theta(o_i |q)}{\pi_{\theta_{\mathrm{old}}}(o_i |q)} A_i,\;
    \\
    & \mathrm{clip}\ \!\bigl(\frac{\pi_\theta(o_i |q)}{\pi_{\theta_{\mathrm{old}}}(o_i |q)},\,1-\varepsilon,\,1+\varepsilon\bigr)\,A_i
    \bigr)
    \\ 
    &- \beta\,\mathbb{D}_{\mathrm{KL}}(\pi_\theta \,\|\, \pi_{\mathrm{ref}})
    \Bigr), \\
    & \mathbb{D}_{\mathrm{KL}}\bigl(\pi_\theta \,\|\, \pi_{\mathrm{ref}}\bigr)
    = \frac{\pi_{\mathrm{ref}}(o_i|q)}{\pi_{\theta}(o_i|q)}
    \\
    & \;-\;\log\!\frac{\pi_{\mathrm{ref}}(o_i|q)}{\pi_{\theta}(o_i|q)}
        \;-\;1,
    \end{split}
\end{equation} 
where $\varepsilon$ and $\beta$ are hyperparameters, and $A_i$ represents the advantage, calculated using rewards $\{r_1, r_2, \dots, r_G\}$ corresponding to outputs within each group:
\begin{equation}
    A_i = \frac{r_i - \mathrm{mean}(\{r_1, r_2, \dots, r_G\})}
           {\mathrm{std}(\{r_1, r_2, \dots, r_G\})}.
\end{equation}

\input{table_beir_sft_grpo}
\input{table_case_study}

\section{BEIR Benchmark Detail}
\label{sec:appendix_beir}
In this paper, we adopt 14 datasets from the BEIR benchmark \citep{thakur2021beir}: SciFact (SF), NFCorpus (NFC), TRECCOVID (TC), ArguAna (AA), FiQA2018 (FQA), Quora (QRA), DBPedia (DBP), Touche2020 (TCH), SciDOCS (SD), Fever (FVR), Climate-Fever (CFV), HotpotQA (HQA), MSMARCO-dev (MM), and Natural Questions (NQ).

\section{SFT \textit{vs} GRPO}
\label{sec:appendix_sft_grpo}
Here, we compare 0.5B ProRank with different prompt warmup training: supervised fine-tuning (SFT) and popular GRPO. 
The results are listed in Table \ref{table:beir_sft_grpo}. 
We can find that ProRank with GRPO prompt warmup performs better than the SFT one. It might be attributed to the multiple rewards optimization strategy, making the model more relevant and accurate in the first stage of training, thereby improving the second stage's performance. This also demonstrates the superiority of reinforcement learning in the first stage of training.

\section{Case Study}
\label{sec:appendix_case_study}
Here, we conduct a case study, as shown in Table \ref{tab:case_study}, to intuitively compare the performance of different models on document reranking.

For baselines, we find that \textit{bge-m3} performs poorly on the second document's ranking. Although the second document does not explicitly mention ``Mockingbird'', it should be ranked higher than the third and fourth documents since it directly identifies \textbf{Harper Lee} as the author. 
This might be attributed to limited commonsense understanding of \textit{bge-m3}.
The \textit{bge-gemma} and proprietary model \textit{voyage-rerank-2} rank the fourth document (containing ``Mockingbird'') higher than other models. 
% This is probably because they rely more heavily on lexical matching than on semantic understanding.
This could indicate a greater reliance on lexical overlap rather than fine-grained semantic matches

For the proposed model, while the coarse-grained ProRank assigns a score of 1 (relevant) to both the first and second documents, it is impossible to differentiate their relative relevance. 
The fine-grained version of ProRank addresses this limitation by providing more nuanced relevance scores. 
Notably, the proposed ProRank with fine-grained scores produces rankings that closely align with human judgment, validating its effectiveness.
% Notably, both the proposed fine-grained ProRank and the proprietary \textit{cohere-rerank-3.5} models produce rankings that closely align with human judgment, validating the effectiveness of the proposed two-stage approach even when compared to famous commercial models.

\section{Discussion on Score Granularity}
\label{sec:appendix_granularity}
Here, we deep into score granularity's impact on reranking performance.
Table \ref{table:beir} demonstrates that ProRank with the fine-grained scoring version consistently outperforms the coarse-grained one. 
For 0.5B ProRank, the fine-grained scoring version achieves a $1.02\%$ absolute improvement in average NDCG@10 ($55.57\%$ \textit{vs} $54.55\%$). 
This improvement becomes more significant with 1.5B ProRank, where the fine-grained scoring version delivers a $1.76\%$ improvement ($57.49\%$ \textit{vs} $55.73\%$). 
Table \ref{table:rest_main} further confirms these findings across languages and domains. 
The average improvement from the fine-grained scoring version is $1.16\%$ for 0.5B ProRank and $2.27\%$ for 1.5B ProRank. 
% The benefits are particularly significant for the Covid dataset, showing a remarkable $4.23\%$ improvement ($89.78\%$ \textit{vs} $85.55\%$) with fine-grained 1.5B ProRank.
These results demonstrate that fine-grained scoring effectively improves document reranking quality. It may be because it captures subtle relevance differences and provides more informative training signals, aligning better with human judgment patterns. 

\section{Comparison with Cutting-edge LLMs}
\label{sec:appendix_cutting_edge}
We further evaluate ProRank by comparing it with two powerful LLMs: Gemini and GPT. The results, presented in Table \ref{tab:cutting_edge}, reveal a significant advantage for ProRank in pointwise document reranking. 
For instance, on SciFact, ProRank 0.5B surpasses Gemini and GPT by $9.18\%$ and $5.77\%$, respectively; the larger ProRank 1.5B achieves even greater improvements of $12.18\%$ and $8.77\%$. 
Notably, ProRank attains these results in mere minutes, compared to the hours required by the LLMs. 
This combination of superior effectiveness and drastically lower computational cost not only highlights ProRank's superiority for reranking but also confirms its strong practical applicability.

\begin{table}[htbp]
\begin{tabular}{l r r}
\hline
\textbf{Model} & \textbf{SciFact} & \textbf{Time (s)} \\
\hline
Gemini-2.5-flash & 67.97 & 3:33:59 \\
GPT5.1 & 71.38 & 5:55:18 \\
\hline
ProRank-0.5B & 77.15 & 00:03:01 \\
ProRank-1.5B & 80.15 & 00:06:17 \\
\hline
\end{tabular}
\caption{Results of Gemini, GPT, and ProRank on the SciFact dataset. NDCG@10 serves as the metric.}
\label{tab:cutting_edge}
\end{table}

\input{table_efficiency}

\section{Discussion on Efficiency}
\label{sec:appendix_efficiency}
Following the scaling law \citep{kaplan2020scaling}, smaller models are generally more efficient. We quantitatively compare the efficiency of different-scale LLM rerankers, as detailed in Table \ref{tab:latency}. The results confirm that our smaller-scale ProRank is significantly more efficient than baseline rerankers using larger-scale LLMs.

%% file: table_beir_sft_grpo.tex
\begin{table*}[htbp]
\setlength\tabcolsep{2pt}
\small
\centering
\begin{threeparttable}
\caption{Results of ProRank with SFT and GRPO prompt warmup on BEIR benchmark, with NDCG@10 as the main metric. ).
}
\label{table:beir_sft_grpo}
\begin{tabular}{lcccccccccccccccccc}
\toprule
Model & Params & SF & NFC  & TC & AA & FQA & QRA & DBP & TCH & SD & FVR & CFV & HQA & MM & NQ & Avg. \\
\midrule
\multicolumn{3}{l}{\textbf{ProRank (fine-grained)}} \\
+ SFT & 0.5B & $76.9$ & $36.9$ & $77.51$ & $62.51$ & $41.16$ & $80.75$ & $43.53$ & $36.15$ & $15.47$ & $81.32$ & $27.49$ & $78.92$ & $40.92$ & $58.03$ & $54.11$ \\
+ GRPO & 0.5B & $77.15$ & $37.10$ & $79.06$ & $57.82$ & $41.09$ & $88.61$ & $44.66$ & $34.58$ & $17.02$ & $88.55$ & $33.02$ & $79.14$ & $41.25$ & $58.89$ & $55.57$ \\
\bottomrule
\end{tabular}
\end{threeparttable}
\end{table*}

%% file: table_case_study.tex
\begin{table*}[htbp]
\large
\caption{Case study of different model rankings for query \textit{``Who wrote 'To Kill a Mockingbird'?''} HM: Human; MB: \textit{mxbai-rerank-large-v1}; M3: \textit{bge-rerank-v2-m3}; GEM: \textit{bge-rerank-v2-gemma}; CH: \textit{cohere-rerank-3.5}; VY: \textit{voyage-rerank-2}; PR$_{0.5B}^{C}$: coarse-grained \textit{ProRank-0.5B}; PR$_{1.5B}^{C}$: coarse-grained \textit{ProRank-1.5B}; PR$_{0.5B}^{F}$: fine-grained \textit{ProRank-0.5B}; PR$_{1.5B}^{F}$: fine-grained \textit{ProRank-1.5B}.}
\label{tab:case_study}
\vspace{0.5em}
\resizebox{\textwidth}{!}{%
\begin{tabular}{p{7cm}ccccccccccc}
    \toprule
    \textbf{Document} & 
    \textbf{HM} & 
    \textbf{MB} & 
    \textbf{M3} & 
    \textbf{GEM} & 
    \textbf{CH} & 
    \textbf{VY} & 
    \textbf{PR$_{0.5B}^{C}$} & 
    \textbf{PR$_{1.5B}^{C}$} & 
    \textbf{PR$_{0.5B}^{F}$} &
    \textbf{PR$_{1.5B}^{F}$} \\
    \midrule
    ``To Kill a Mockingbird' is a novel by \textbf{Harper Lee} published in 1960.' & \multirow{2}{*}{0} & \multirow{2}{*}{0} & \multirow{2}{*}{0} & \multirow{2}{*}{0} & \multirow{2}{*}{0} & \multirow{2}{*}{0} & \multirow{2}{*}{1} & \multirow{2}{*}{1} & \multirow{2}{*}{0} & \multirow{2}{*}{0} \\
    \midrule
    \textbf{Harper Lee}, an American novelist, was born in 1926 in Monroeville, Alabama. & \multirow{2}{*}{1} & \multirow{2}{*}{1} & \multirow{2}{*}{2} & \multirow{2}{*}{1} & \multirow{2}{*}{1} & \multirow{2}{*}{1} & \multirow{2}{*}{0} & \multirow{2}{*}{0} & \multirow{2}{*}{1}  & \multirow{2}{*}{1} \\
    \midrule
    The 'Harry Potter' series,  consisting of seven fantasy novels, was written by British author J.K. Rowling. & \multirow{3}{*}{2} & \multirow{3}{*}{3} & \multirow{3}{*}{1} & \multirow{3}{*}{3} & \multirow{3}{*}{2} & \multirow{3}{*}{3} & \multirow{3}{*}{2} & \multirow{3}{*}{2} & \multirow{3}{*}{2} & \multirow{3}{*}{2} \\
    \midrule
    What does a Mockingbird eat? & 3 & 2 & 3 & 2 & 3 & 2 & 3 & 3 & 3 & 3 \\
    \bottomrule
\end{tabular}
}
\end{table*}

%% file: table_efficiency.tex
\begin{table}[htbp]
\center
\small
\begin{tabular}{l r r r}
\hline
\textbf{Model} & \textbf{Params} & \textbf{Load (s)} & \textbf{Avg. (ms)} \\
\hline
bge-v2-gemma  & 2B & 1.26 & 64.0 \\
rankllama-7b     & 7B & 4.88 & 49.7 \\
rankllama-13b     & 13B & 6.77 & 81.0  \\
\hline
ProRank-0.5B           & 0.5B & 1.29 & 27.2  \\
ProRank-1.5B           & 1.5B & 1.55 & 32.7 \\
\hline
\end{tabular}
\caption{Latency of different scales of LLM rerankers. rankllama is from \citep{ma2024fine}. Avg. means average latency.}
\label{tab:latency}
\end{table}